\documentclass[nolinenumbers]{aastex631}

\shorttitle{Global Coronal Equilibria with Solar Wind Outflow}
\shortauthors{Rice and Yeates}

\graphicspath{{./}{figures/}}

\usepackage{amsmath}

\begin{document}

\title{Global Coronal Equilibria with Solar Wind Outflow}

\author{Oliver E. K. Rice}
\author{Anthony R. Yeates}
\affiliation{Department of Mathematical Sciences, Durham University, Durham, DH1 3LE, UK \\}

\submitjournal{The Astrophysical Journal}
\accepted{1st October 2021}

\begin{abstract}
Given a known radial magnetic field distribution on the Sun's photospheric surface, there exist well-established methods for computing a potential magnetic field in the corona above. Such potential fields are routinely used as input to solar wind models, and  to initialize magneto-frictional or full magnetohydrodynamic simulations of the coronal and heliospheric magnetic fields. We describe an improved magnetic field model which calculates a magneto-frictional equilibrium with an imposed solar wind profile (which can be Parker's solar wind solution, or any reasonable equivalent). These `outflow fields' appear to approximate the real coronal magnetic field more closely than a potential field, take a similar time to compute, and avoid the need to impose an artificial source surface. Thus they provide a practical alternative to the potential field model for initializing time-evolving simulations or modeling the heliospheric magnetic field. We give an open-source Python  implementation in spherical coordinates and apply the model to data from Solar Cycle 24. The outflow tends to increase the open magnetic flux compared to the potential field model, reducing the well known discrepancy with in situ observations.
\end{abstract}


\section{Introduction}

The magnetic structure of the Sun's corona has historically been difficult to model. A variety of methods have been used, ranging from simple potential-field models to full magnetohydrodynamic simulations \citep[see the reviews by][]{2012LRSP....9....6M, 2018LRSP...15....4G}.  One of the most popular and well-established models is the Potential Field Source-Surface (or PFSS) model, first used by \citet{1969SoPh....9..131A} and \citet{1969SoPh....6..442S}. This seeks to calculate a potential magnetic field that satisfies a lower boundary condition at the base of the corona, usually provided by magnetogram data. Such a magnetic field is current-free and takes the form 
\begin{equation}
    {\bf B} = \nabla \Phi,
\end{equation}
for some scalar field $\Phi$. Combined with the solenoidal condition $\nabla \cdot \bf{B} = 0$, we find that $\Phi$ is a solution to Laplace's equation.

There are many advantages to using the PFSS model, not least that it is comparatively simple to compute and requires less boundary data compared to more physically accurate models. In fact, despite the enormous increases in computing capability since their inception, PFSS solutions of the corona are still widely used for a number of applications, albeit with various additions and modifications to the original models \citep[e.g.,][]{2020ApJS..246...23B, https://doi.org/10.1029/2001JA007550}. In particular, models based on a PFSS framework are among the most widely used bases for space weather prediction \citep{2018SpWea..16.1644M}.

In order to calculate a potential field, an upper boundary condition must be chosen. PFSS models specify that the magnetic field lines are purely radial at a given radius, often taken to be $2.5R_\odot$. The justification for this is that the solar wind opens out the potential arcades such that they become radial at around this altitude. However, there are significant discrepancies between this assumption and physical observations of the corona. Various attempts have been made to address this problem, such as experimenting with a non-spherical source surface \citep[e.g.,][]{1982SoPh...77..363L}, or allowing the source-surface height to vary over time \citep{2020ApJ...889L..28V}. The latter authors showed that the PFSS open magnetic flux derived from two magnetogram datasets can match observations at 1AU, but this can not be used for prediction since the optimum source-surface radius varies in an irregular and unpredictable manner. In general, the predicted amount of open magnetic flux in PFSS models does not match observations, both at 1AU and closer to the Sun. This is the so-called `open-flux problem' \citep{2017ApJ...848...70L}, whereby the heliospheric magnetic flux is often measured at twice or more than the value predicted by PFSS models. There still a considerable difference in the open flux even quite close to the Sun, as measured by the Parker Solar Probe \citep{2021A&A...650A..18B}, and the precise origin of this discrepancy remains unclear \citep{2020JGRA..12526005V}.

We also consider that the corona is not everywhere current-free, as is assumed in potential field models. In the lower corona small-scale structures are far from current-free, but in general these do not greatly affect the global magnetic structure. However at higher altitudes the solar wind can induce currents that can greatly affect the structure even in a steady equilibrium. This is not captured accurately by PFSS models and leads to unrealistic streamer shapes in the upper corona  when compared to eclipse observations or full MHD (magnetohydrodynamic) simulations \citep{2006ApJ...653.1510R}.

As the most widely-used alternative to PFSS models, full MHD codes  solve the magnetic and fluid equations, and a realistic stationary magnetic field can be found by allowing the simulation to relax to an equilibrium. Compared to the PFSS model, MHD codes are very expensive to run and require more boundary data than line-of-sight observations can provide. \citet{2018LRSP...15....4G} discusses the history of MHD applications, which have only relatively recently been able to accurately model the corona in full spherical coordinates, using realistic lower boundary data \citep[e.g.,][]{doi:10.1063/1.873474,1993SoPh..146..377U}. The coronal topology of these models compares favourably with PFSS equivalents, and more realistic streamer shapes are observed. Thus, despite MHD models being too expensive for many applications, they can be used as a more reliable reference point to test the accuracy of other variations on the PFSS model. For example, \citet{1999JGR...104.9809L} modeled the sun using MHD simulations for a whole month during 1996, and found that the simulated coronal structures agreed very closely to images collected from ground-based telescopes and spacecraft. With appropriate thermodynamics, more recent developments of the model are found to produce coronae that agree well with eclipse observations \citep{2018NatAs...2..913M}. The MHD scheme developed originally by \citet{1999JCoPh.154..284P} has also been extensively used in space weather predictions, being used to model specific events such as coronal mass ejections, which can not be captured in a PFSS or similar model. Other applications of MHD to space weather predictions include the EUHFORIA project \citep{2018JSWSC...8A..35P}.

The aim of this paper is to improve the accuracy of the PFSS model without the significant added computational expense and additional boundary data required by full MHD models. Specifically, we seek to include the effect of a solar wind outflow near the upper boundary. For simplicity, in our model we must assume that the solar wind is radial and has only radial dependence. This is quite a restrictive assumption given the solar wind speed does vary depending on latitude. In future it may be possible to generalise the method to account for more realistic solar wind flows. One approach would be to solve directly for a steady MHD equilibrium that satisfies the full MHD equations with the time derivative terms removed. For an axisymmetric helmet streamer, this was done by \citet{1971SoPh...18..258P} using an iterative numerical approach. Recently, \citet{2020SoPh..295..145W} have developed an optimization method for finding such equilibria in three dimensions. This is less computationally expensive than the standard method of allowing full MHD simulations to relax, but is still more complex and less mathematically well-defined than the PFSS model. 

Our work is motivated by another simplified approach that has been developed for global modelling of the solar coronal magnetic field: the magneto-frictional (MF) model \citep[e.g.,][]{2012LRSP....9....6M}. On a global scale, this technique has been applied with time-dependent lower boundary conditions from either surface flux transport models \citep[e.g.,][]{2008SoPh..247..103Y,2014SoPh..289..631Y} or more direct assimilation of magnetogram data \citep{2016ApJ...823...55W, 2020ApJS..250...28H}. In the MF model (more details in Section \ref{sec:approach}), the fluid equations in a full MHD system are removed and in their place the system is closed by specifying an explicit form of the plasma velocity in terms of the magnetic field, along with an additional radial outflow to represent the effect of the solar wind. It has been observed \citep{2010JGRA..115.9112Y} that when using this model the open flux at the top boundary increases compared to a potential field, and indeed when left to run the system will relax to a new equilibrium in a day or so, taking into account the effect of the solar wind. The price to pay for MF being a purely magnetic model is that the  radial outflow is simply an imposed function, unlike in the more physically complete model of full MHD which includes density and temperature so as to determine a self-consistent velocity field.

It is important to note that the currents in the upper corona caused by the solar wind do not appear to have any effect on the behaviour of the small-scale structures (e.g. flux ropes) near the solar surface \citep{2009ApJ...699.1024Y} and so the currents in the two regions can be regarded as independent. The idea in this paper is to account for the currents in the upper corona, while neglecting those in the low corona whose origin is more difficult to capture. By neglecting these currents, when modelling the lower corona our method does not provide any improvement over standard PFSS models. If this region is of interest an alternative approach (either MF or MHD) that accounts for time-dependent driving at the photosphere should be used.

The equilibrium solutions of these MF models show qualitatively more realistic streamer shapes than a PFSS field \citep[for example, Figure 10 of][]{2012LRSP....9....6M}. But again, the MF model requires time integration to reach a steady equilibrium. In this paper, we aim to calculate equilibrium solutions of the MF model, without the need for time evolution. The calculation is direct and does not require any optimization techniques, such as those used in \citet{2020SoPh..295..145W}, and with some refinement should be similarly cheap as traditional PFSS models. We show that these `outflow' solutions appear to exhibit more accurate streamer shapes than PFSS models and avoid an unrealistic boundary layer at $2.5R_{\odot}$. They also result in higher open flux than PFSS equivalents, which perhaps could provide a partial solution to the open flux problem.

We begin in Section \ref{sec:approach} by briefly describing the MF model. When the solar wind is not taken into account, potential fields are equilibrium solutions to this model, but we generalize the long-standing method of finding potential solutions to allow for the currents generated in the upper corona by the outflow. The outflow velocity is required to be radial, and a function of radius only. An approximation to the original Parker solar wind solution \citep{1958ApJ...128..664P} is the example we use in this paper, but any suitable function will suffice. A modified eigenfunction expansion provides an equilibrium solution to this model, similarly to the method used by most  PFSS codes. However, there are various numerical challenges that must be overcome, which we describe in Section \ref{sec:num} along with their resolutions. The result is a numerical method designed for a staggered grid in modified spherical coordinates, although it can easily be adapted for other coordinate systems. The resulting magnetic field is divergence-free to machine precision on this staggered grid.

The code can calculate PFSS fields by setting the outflow velocity to zero, as illustrated in Section \ref{sec:pf}. This allows us to easily compare the outflow field and potential field for a series of magnetograms, measured throughout Solar Cycle 24 (Section \ref{sec:cycle24}). We observe a significant change in the shape of streamers and the structure of the upper corona, such that the structures appear to be more similar to MHD solutions and eclipse observations than an equivalent PFSS solution. There is also a significant increase in the open flux measured at $2.5R_{\odot}$. This goes some way towards addressing the discrepancies discussed by \citet{2017ApJ...848...70L}.

\section{Modelling approach} \label{sec:approach}

\subsection{Magneto-frictional equilibria with outflow}

In magnetohydrodynamic models, the velocity field is determined by the momentum equation
\begin{equation}
\rho\frac{\mathrm{D}{\bf v}}{\mathrm{D}t} = \frac{1}{\mu_0}(\nabla\times{\bf B})\times{\bf B} - \nabla p - \rho\nabla\Psi,
\end{equation}
coupled to the ideal induction equation
\begin{equation}
\frac{\partial{\bf B}}{\partial t} = \nabla\times({\bf v}\times{\bf B}),
\end{equation}
along with additional fluid equations to close the system. In the magneto-frictional method, pressure gradients and gravity are neglected, and instead a frictional velocity is imposed as
\begin{equation}
\nu{\bf v}= (\nabla\times{\bf B})\times{\bf B},
\label{eqn:vxb}
\end{equation}
so that the induction equation leads to monotonic relaxation towards a stationary force-free field with $(\nabla\times{\bf B})\times{\bf B}={\bf 0}$. The friction coefficient $\nu$ is typically given the form $\nu=\nu_0|{\bf B}|^2$ (with some minimum value imposed) so that the overall evolution is independent of the magnitude of ${\bf B}$ and relaxation is not unduly slow near to magnetic null points \citep{1986ApJ...309..383Y}.

In the outer corona, the solar wind outflow prevents the magnetic field from being force-free, but this effect can be approximated in the magneto-frictional model by relaxing towards an equilibrium with a specified outflow ${\bf v}_{\rm out}$, thus choosing ${\bf v}$ according to
\begin{equation}
\bf v = \frac{(\nabla\times{\bf B})\times{\bf B})}{\nu} + {\bf v}_{\rm out}
\end{equation} 
This \textit{ad hoc} approach was introduced by \citet{2006ApJ...641..577M}, and has subsequently been used in global magneto-frictional models of solar and stellar coronae \citep[e.g.][]{2014SoPh..289..631Y, 2016MNRAS.456.3624G, 2018ApJ...869...62M, 2020SoPh..295..101M}. When a potential field is chosen to initialize the model, there is an initial period of up to a few days' evolution during which the system adjusts itself into the new equilibrium state.
In this paper, we propose to solve directly for equilibria of this model, avoiding this initial unphysical period of adjustment.

It is critical to note that the equilibria are not given by ${\bf v}={\bf 0}$ but rather by ${\bf v}\times{\bf B} = {\bf 0}$, thanks to the form of the induction equation.  Thus to calculate such an equilibria directly, the equation we need to solve is
\begin{equation}
\left(\frac{(\nabla\times{\bf B})\times{\bf B}}{\nu} + {\bf v}_{\rm out}\right)\times{\bf B} ={\bf 0},
\label{eqn:eqm}
\end{equation}
for specified ${\bf v}_{\rm out}$. For simplicity, we will assume that the outflow velocity is purely radial, depends only on radius, and is constant in time, so ${\bf v}_{\rm out} = v_{\rm out}(r){\bf e}_r$. In reality the solar wind speed does vary with latitude, and it may be possible to generalise our method to take this into account in future. We choose the wind speed in general to match Parker's solar wind solution \citep{1958ApJ...128..664P}, which for altitudes below the critical radius $r_c$ (around $10 R_\odot$ for typical coronal temperatures) is approximately
\begin{equation}
v_{\rm out}(r) = v_1\frac{r_1^2\mathrm{e}^{-2r_c/r}}{r^2\mathrm{e}^{-2r_c/r_1}}.
\end{equation} 
This function follows from the well-known implicit equation for the solar wind
\begin{equation}
    \left(\frac{v(r)}{v_c}\right)^2 - 2\ln\left(\frac{v(r)}{v_c}\right) = 4\ln\left(\frac{r}{r_c}\right) + \frac{4r_c}{r} - 3,
\end{equation}
after neglecting the $v^2$ term, and scaling as appropriate so that $v(r) = v_1$ at the top boundary $r = r_1$. This approximation matches the exact solution of the implicit equation very closely throughout the computational domain if $r_1 = 2.5R_{\odot}$, but not at altitudes significantly higher than this. Thus in order to calculate accurate fields higher in the corona a more realistic solar wind approximation must be used. 

The shape of a solution to Equation \eqref{eqn:eqm} is determined by the product $\nu_0 v_1$, such that altering the value of $\nu_0$ is equivalent to scaling the outflow velocity function by a constant. We can take the flow speed $v_1$ from the Parker solution, which for an isothermal corona at $2\,\mathrm{M}\mathrm{K}$ gives $v_1\approx 157\,\mathrm{km}\,\mathrm{s}^{-1}$ at $r_1=2.5R_\odot$. However, it is difficult to determine an \textit{a priori} value for the constant $\nu_0$, as it does not directly correspond to a physical quantity. Previous magneto-frictional simulations of the global corona  have used $\nu_0$ values of the order $\nu_0\sim 5\times 10^{-17}\,\mathrm{s}\,\mathrm{cm^{-2}}$ \citep[cf.][]{2016A&A...594A..98Y}, so we will adopt this value for our computations in this paper.

\subsection{Solution technique in spherical geometry}

Since our imposed ${\bf v}_{\rm out}$ depends only on the radial coordinate $r$, the basic idea is to look for solutions to \eqref{eqn:eqm} of the form
\begin{equation}
\textbf{B} = f(r) \nabla X, \label{beq}
\end{equation}
where $f(r)$ and $X(r, \theta,\phi)$ are functions to be determined. Thus a potential field would correspond to the special case of constant $f$. It follows that
\begin{eqnarray}
[(\nabla\times{\bf B})\times{\bf B}]\times{\bf B} = -({\bf B}\cdot\nabla X)(\nabla f\times{\bf B}) = -\frac{|{\bf B}|^2f'}{f}{\bf e}_r\times{\bf B}.
\end{eqnarray}

Substituting this into the equilibrium equation \eqref{eqn:eqm} with ${\bf v}_{\rm out} = v_{\rm out}(r){\bf e}_r$ and $\nu = \nu_0|{\bf B}|^2$ reduces to the ordinary differential equation
\begin{equation}
    f'(r) = \nu_0 v_{\rm out}(r)f(r).
    \label{eqn:fode}
\end{equation}
(In our implementation, this equation will be slightly modified due to the use of a stretched radial coordinate as described in Section \ref{sec:eig} below.) Notice that, in the absence of outflow ($v_{\rm out}=0$), Equation \eqref{eqn:fode} gives $f=\textrm{constant}$, corresponding to a potential field.  The function $X$ is then determined by the solenoidal condition $\nabla\cdot{\bf B}=0$, which gives the partial differential equation
\begin{eqnarray}
f\Delta X + \nabla f\cdot\nabla X = 0. 
\end{eqnarray}
Again, when $f$ is constant, this reduces to the usual Laplace equation $\Delta X=0$ for a potential field. Eliminating $f$ with \eqref{eqn:fode} gives
\begin{equation}
\Delta X + \nu_0{\bf v}_{\rm out}\cdot\nabla X = 0.    \label{eqn:Xpde}
\end{equation}

\subsection{Eigenfunction expansion} \label{sec:eig}

Since equation \eqref{eqn:Xpde} is linear, we seek to write $X$ in terms of eigenmodes. Our numerical implementation in Section \ref{sec:num} uses a grid equally spaced in stretched spherical coordinates $(\rho, s, \phi)$ satisfying
\begin{equation} 
\rho = \ln(r), \quad s =\cos \theta, \quad \phi = \phi,
\end{equation}
rather than normal spherical coordinates $(r,\theta,\phi)$. Thus it is convenient to derive the equations in these coordinates. The coordinate scale factors in this system are
\begin{equation}
h_\rho = r = e^\rho, \,\,\, h_s = \frac{r}{\sin \theta} = \frac{e^\rho}{\sqrt{1-s^2}}, \,\,\, h_\phi = r\sin \theta = e^\rho \sqrt{1-s^2}. \label{hfacts}
\end{equation}
Writing the unknown function as an eigenfunction expansion
\begin{equation}
    X(\rho, s, \phi) = \sum_{l,m}C_{l,m}R_l(\rho)Q_{l,m}(s)\Phi_m(\phi),
\end{equation}
we substitute into \eqref{eqn:Xpde} to obtain the three eigenfunction equations
\begin{align}
   R_l'' + (1 + \nu_0\mathrm{e}^\rho v_{\rm out})R_l' &= l(l+1)R_l, \label{eqn:rmode}\\
   (1-s^2)Q_{l,m}'' - 2sQ_{l,m}' + l(l+1)Q_{l,m} &= \frac{m^2}{1-s^2}Q_{l,m},\label{eqn:smode}\\
   \Phi_m'' &= - m^2\Phi_m,\label{eqn:pmode}
\end{align}
where $l$ and $m$ are integers with $-l\leq m \leq l$. The latitudinal and azimuthal equations are the same as for the Laplace equation $\Delta X=0$, yielding the associated Legendre polynomials $Q_{l,m}$ and trigonometric functions $\Phi_m$ that are familiar from the potential field model. However, the radial eigenfunctions differ from a potential field due to the presence of $v_{\rm out}$ term, and we must additionally solve for the function $f(r)$, using the equation
\begin{equation}
f'(\rho) = e^\rho v_{out}(\rho)f(\rho).
\end{equation}
Notice that this differs to \eqref{eqn:fode} as the coordinate in the radial direction has been stretched.

In the potential field case where $v_{\rm out}=0$, equation \eqref{eqn:rmode} has the exact general solution
\begin{equation}
    R_l(\rho) = A\mathrm{e}^{l\rho} + B\mathrm{e}^{(-l-1)\rho},
\end{equation}
but in the presence of outflow, the equation must be solved numerically. In practice, we find that although we can solve \eqref{eqn:rmode} for $R_l$, solving for the radial function $f$ is numerically unstable as the value for $f$ at the top boundary is far too large to compute. The solution to this problem is to solve for the rescaled eigenfunctions
\begin{equation}
H_l(\rho) = \mathrm{e}^{-\rho}f(\rho)R_l(\rho),
\label{eqn:hp}
\end{equation}
which may be shown using \eqref{eqn:rmode} and \eqref{eqn:fode} to satisfy
\begin{equation}
H_l'' + (3-\nu_0\mathrm{e}^\rho v_{\rm out})H_l' - \Big[l(l+1) - 2 + 3\nu_0\mathrm{e}^\rho v_{\rm out} +  \nu_0\mathrm{e}^\rho v_{\rm out}'\Big]H_l = 0.
\label{eqn:hmode}
\end{equation}

In terms of the eigenfunctions, ${\bf B}$ has the form
\begin{align}
    {\bf B} &= \sum_{l,m}C_{l,m}\left[\frac{fR_l'}{\mathrm{e}^\rho}Q_{l,m}\Phi_m{\bf e}_\rho + \frac{fR_l}{\mathrm{e}^\rho}\sqrt{1-s^2}Q_{l,m}'\Phi_m{\bf e}_s + \frac{fR_l}{\mathrm{e}^\rho}\frac{Q_{l,m}}{\sqrt{1-s^2}}\Phi_m'{\bf e}_\phi\right]\\
    &= \sum_{l,m}C_{l,m} \left[G_lQ_{l,m}\Phi_m{\bf e}_\rho + H_l\sqrt{1-s^2}Q_{l,m}\Phi_m{\bf e}_s + H_l\frac{Q_{l,m}}{\sqrt{1-s^2}}\Phi_m'{\bf e}_\phi\right],\label{eqn:beig}
\end{align}
where $H_l$ are the rescaled eigenfunctions in \eqref{eqn:hp} and we define the combination
\begin{eqnarray}
G_l(\rho) = \frac{f(\rho)R_l'(\rho)}{\mathrm{e}^\rho}. \label{eqn:g}
\end{eqnarray}
Thus ${\bf B}$ may be calculated in a similar way to the classical potential field, by solving the eigenfunction equations \eqref{eqn:smode}, \eqref{eqn:pmode} and \eqref{eqn:hmode}. As in the potential field model, the coefficients $C_{l,m}$ are determined by matching the observed radial field distribution $B_\rho(\rho_0,s,\phi)$ on the lower boundary $\rho=
\rho_0$. 
The corresponding lower boundary condition for $H_l$ is determined by choosing $G_l(\rho_0)=1$. This leads to quite a complex boundary condition, but a good approximation can be used instead by making the assumption that $v'_{\rm out}(\rho_0) = 0$ (which is very nearly true for a realistic outflow function). With this assumption, it follows that a suitable lower boundary condition is 
\begin{equation}
(H_l(\rho)e^\rho)'\vert_{\rho_0} = e^{\rho_0}.
\label{eqn:orbnd}
\end{equation}
A numerical approximation of this is used in the code.

As the required second boundary condition for $H_l$, we set $H_l(\rho_1)=0$ at some outer boundary $\rho=\rho_1$, so that ${\bf B}$ is purely radial there. Provided $\rho_1$ is high enough, this condition does not have a significant influence on the shape of the magnetic field, since the field lines tend to be radial already in the upper part of the domain when outflow is present. This is in contrast to the potential field where the radial field condition at the source surface has a significant effect on the shape of the field.

\section{Numerical Implementation} \label{sec:num}

We have written a numerical code to calculate the outflow equilibria in spherical geometry. This Python code is open source and freely available at \url{https://github.com/oekrice/outflow}. Our approach is to calculate ${\bf B}$ on a staggered grid \citep{1966ITAP...14..302Y}, such that $\nabla\cdot{\bf B}=0$ to machine precision in a particular discretization. This will make our solutions suitable for initializing future magneto-frictional simulations using our numerical code on the same grid \citep[e.g.,][]{2021SoPh..296..109B}. Imposing the discrete solenoidal condition on a finite grid leads to discrete eigenfunctions $H_l$, $Q_{l,m}$ and $\Phi_m$ that are only approximations to the exact analytical eigenfunctions, and to eigenvalues $m$ and $l$ that are no longer necessarily integers. A similar approach was used by \citet{2000ApJ...539..983V} for potential fields, and implemented in the Python potential field solver of \citet{2020JOSS....5.2732S} that uses the same $(\rho, s, \phi)$ grid as in this paper. In the following subsections, we describe the numerical method in more detail. The same approach could be implemented on other grids, including cartesian coordinates, by modifying the geometrical factors.

\subsection{Staggered grid}

We number the cells with $i,j,k$ indices for the $\rho,s,\phi$ directions respectively. The indices take integer values at the grid points, which are equally spaced in the $(\rho,s,\phi)$ coordinates and given by
\begin{eqnarray}
\rho^i &=& \rho_0 + i\delta\rho, \quad \delta\rho = (\rho_1-\rho_0)/n_\rho,\\
s^j &=& -1 + j\delta s, \quad \delta s = 2/n_s,\\
\phi^k &=& k\delta\phi, \quad \delta\phi = 2\pi/n_\phi.
\end{eqnarray}
The $i,j,k$ indices take half-integer values at the cell faces, whose areas 
may be calculated from the coordinate transform by integration of the coordinate scale factors \eqref{hfacts}. This gives
\begin{eqnarray}
S_\rho^{i,j+\frac12,k+\frac12} &=&e^{2\rho^i} \delta s \delta \phi \\
S_s^{i+\frac12,j,k+\frac12} &=&\frac{1}{2}(e^{2\rho^{i+1}} - e^{2\rho^i})\sigma^j \delta \phi \\
S_\phi^{i+\frac12,j+\frac12,k} &=&\frac{1}{2}(e^{2\rho^{i+1}} - e^{2\rho^j})(\arcsin(s^{j+1})-\arcsin(s^j)), 
\end{eqnarray}
where $\sigma^j = \sqrt{1-(s^{j})^2}$ is a quantity that appears frequently.
The magnetic field components are defined on the corresponding faces and denoted $B_\rho^{i,j+\frac12,k+\frac12}$, $B_s^{i+\frac12,j,k+\frac12}$, $B_\phi^{i+\frac12,j+\frac12,k}$. 

The magnetic field is expanded in a finite series of discrete eigenfunctions, so that analogously to \eqref{eqn:beig} we have
\begin{align}
B_\rho^{i,j+\frac12,k+\frac12} &= \sum_{l,m} C_{l,m} G_l^i Q_{l,m}^{j+\frac12}\Phi_m^{k+\frac12},\label{eqn:dbr}\\
B_s^{i+\frac12,j,k+\frac12} &= \sum_{l,m} C_{l,m} H_l^{i+\frac12}\sigma^j(Q_{l,m}')^j\Phi_m^{k+\frac12},\label{eqn:dbs}\\
B_\phi^{i+\frac12,j+\frac12,k} &= \sum_{l,m} C_{l,m} H_l^{i+\frac12}\frac{1}{\sigma^{j+\frac12}}Q_{l,m}^{j+\frac12}(\Phi'_m)^k.\label{eqn:dbp}
\end{align}

\subsection{Discrete solenoidal condition}

On each grid cell, we impose the solenoidal condition in integral form, which requires
\begin{multline}
B_\rho^{i+1,j+\frac12,k+\frac12}S_\rho^{i+1,j+\frac12,k+\frac12} - B_\rho^{i,j+\frac12,k+\frac12}S_\rho^{i,j+\frac12,k+\frac12} + \\
B_s^{i+\frac12,j+1,k+\frac12}S_s^{i+\frac12,j+1,k+\frac12} -  B_s^{i+\frac12,j,k+\frac12}S_s^{i+\frac12,j,k+\frac12} + \\
B_\phi^{i+\frac12,j+\frac12,k+1}S_\phi^{i+\frac12,j+\frac12,k+1} - B_\phi^{i+\frac12,j+\frac12,k}S_\phi^{i+\frac12,j+\frac12,k} =0.
\label{eqn:sol}
\end{multline}
This translates into equations for $G_l^i$, $Q_{l,m}^{j+\frac12}$ and $\Phi_m^{k+\frac12}$, as follows. Substituting the discrete expansions \eqref{eqn:dbr}, \eqref{eqn:dbs}, \eqref{eqn:dbp} into this condition leads -- for a single mode $l$, $m$ -- to the equation
\begin{multline}
(G_l^{i+1}\mathrm{e}^{2\rho^{i+1}} - G_l^i\mathrm{e}^{2\rho^i})Q_{l,m}^{j+\frac12}\Phi_m^{k+\frac12} \delta s \delta \phi  
+\frac{1}{2} (\mathrm{e}^{2\rho^{i+1}} - \mathrm{e}^{2\rho^i}) H_l^{i+\frac12}\Phi_m^{k+\frac12}  \delta \phi [(Q'_{l,m})^{j+1}(\sigma^{j+1})^2- (Q'_{l,m})^j(\sigma^j)^2] +\\ 
 \frac{1}{2\sigma^{j+\frac12}}H_l^{i+\frac12} Q_{l,m}^{j+\frac12}(\mathrm{e}^{2\rho^{i+1}} - \mathrm{e}^{2\rho^i})[ \arcsin(s^{j+1})-\arcsin(s^j)][(\Phi_m')^{k+1 } - (\Phi_m')^{k}] = 0. \label{eqn:big}
\end{multline}
Since the $(\Phi_m)^{k+\frac12}$ approximate trigonometric functions, we assume the discrete approximation
\begin{eqnarray}
\Phi_m'^{k+1} - \Phi_m'^k = -m^2\Phi_m^{k+\frac12}\delta\phi
\label{eqn:ptrig}
\end{eqnarray}
for some $m$ that would be an integer in the limit $\delta\phi\to 0$ but not necessarily so at our finite resolution. This removes the azimuthal dependence from \eqref{eqn:big} and reduces it to the separable form
\begin{multline}
\frac{2}{H_l^{i + \frac12}}\frac{G_l^{i+1}\mathrm{e}^{2\rho^{i+1}} - G_l^i\mathrm{e}^{2\rho^i}}{\mathrm{e}^{2\rho^{i+1}} - \mathrm{e}^{2\rho^i}} + \\
\frac{1}{Q_{l,m}^{j+\frac12} \delta s}\{(Q_{l,m}')^{j+1}(\sigma^{j+1})^2 - (Q_{l,m}')^j(\sigma^j)^2
- \frac{m^2}{\sigma^{j+\frac12}} Q_{l,m}^{j+\frac12}[ \arcsin(s^{j+1})-\arcsin(s^j)]\} = 0.
\end{multline}
In order that $Q_{l,m}$ approximate the analytical associated Legendre polynomials, we choose a separation constant of the form $l(l+1)$ so that we obtain the radial equation
\begin{equation}
    \frac{2}{H_l^{i + \frac12}}\frac{G_l^{i+1}\mathrm{e}^{2\rho^{i+1}} - G_l^i\mathrm{e}^{2\rho^i}}{\mathrm{e}^{2\rho^{i+1}} - \mathrm{e}^{2\rho^i}} =l(l+1) \label{eqn:gnum}
\end{equation}
and the latitudinal equation
\begin{equation}
    (Q_{l,m}')^{j+1}(\sigma^{j+1})^2 - (Q_{l,m}')^j(\sigma^j)^2
- \frac{m^2}{\sigma^{j+\frac12}} Q_{l,m}^{j+\frac12}[ \arcsin(s^{j+1})-\arcsin(s^j)] = -l(l+1)Q_{l,m}^{j+\frac12}\delta s. \label{eqn:qnum}
\end{equation}
Again, on our finite resolution grid, the $l$ will no longer be precisely integers.

\subsection{Calculation of azimuthal eigenfunctions}

Approximating $(\Phi_m')^k$ by central differences reduces \eqref{eqn:ptrig} to the tridiagonal eigenvalue problem
\begin{equation}
    -\Phi_m^{k+\frac32} + 2\Phi_m^{k+\frac12} - \Phi_m^{k-\frac12} = \lambda_m\Phi_m^{k+\frac12},
\end{equation}
which determines both the eigenfunctions and the values of $m$, from the eigenvalues $\lambda_m = m^2\delta\phi^2$. To ensure periodicity in the azimuthal direction, we need to ensure that the eigenfunctions approximate cosine or sine functions with integer coefficients. The boundary conditions for cosine functions are $\Phi_m^{-\frac12} = \Phi_m^{\frac12}$ and $\Phi_m^{n_\phi+\frac12}=\Phi_m^{n_\phi-\frac12}$. The boundary conditions for sine functions are $\Phi_m^{-\frac12} = -\Phi_m^{\frac12}$ and $\Phi_m^{n_\phi+\frac12}=-\Phi_m^{n_\phi-\frac12}$. By avoiding implementing the periodic boundary conditions directly, we retain a tridiagonal eigenvalue problem that is efficiently solved with a standard solver.

This unusual approach is used instead of a Fast Fourier Transform as we need to have some flexibility in the numerical scheme for the radial eigenfunctions. These radial functions then specify exactly the necessary schemes used in other directions so as to preserve \eqref{eqn:sol}. In contrast, in most PFSS (potential field) codes \citep[e.g.,][]{2011ApJ...732..102T, 2000ApJ...539..983V} the azimuthal eigenfunctions are calculated using a Fourier Transform and the numerical schemes in the radial and latitudinal directions follow from this.

\subsection{Calculation of latitudinal eigenfunctions}

The eigenfunctions $Q_{l,m}^{j+\frac12}$ and possible values of $l$ are determined by \eqref{eqn:qnum}, which is a discrete approximation to the associated Legendre equation \eqref{eqn:smode}. To see this, observe that
\begin{equation}
    \frac{(Q_{l,m}')^{j+1}(\sigma^{j+1})^2 - (Q_{l,m}')^j(\sigma^j)^2}{\delta s} \approx \frac{\mathrm{d}}{\mathrm{d}s}[Q_{l,m}'(1-s^2)],
\end{equation}
and that
\begin{equation}
    \frac{m^2}{\sigma^{j+\frac12}} Q_{l,m}^{j+\frac12}[ \arcsin(s^{j+1})-\arcsin(s^j)] \approx \frac{m^2 Q_{l,m}}{\sqrt{1-s^2}}\frac{\mathrm{d}}{\mathrm{d}s}\textrm{arcsin}(s) = \frac{m^2 Q_{l,m}}{1-s^2}. 
\end{equation}
Approximating the derivatives $(Q_{l,m}')^j$ by central differences $(Q_{l,m}')^j = (Q_{l,m}^{j+\frac12} - Q_{l,m}^{j-\frac12})/\delta s$ reduces 
\eqref{eqn:qnum} to a tridiagonal eigenvalue problem for each $m$ \citep[cf.][]{2000ApJ...539..983V}. Specifically,
\begin{multline}
Q_{l,m}^{j+\frac32}(\sigma^{j+1})^2  -
Q_{l,m}^{j+\frac12}\left[(\sigma^{j+1})^2 + (\sigma^j)^2 + \frac{m^2}{\sigma^{j+\frac12}}[ \arcsin(s^{j+1})-\arcsin(s^j)]\right] + Q_{l,m}^{j-\frac12}(\sigma^j)^2= 
\mu_{l,m} Q_{l,m}^{j+\frac12},
\end{multline}
where $\mu_{l,m} = -l(l+1)\delta s$. The eigenvalues $l$ are different for each eigenvalue $m$, and like $m$ they are approximately integers for small $l$, converging to integers for larger and larger $l$ as $\delta s\to 0$.  In this limit the discrete eigenfunctions $Q_{l,m}^{j+\frac12}$ converge to the associated Legendre polynomials.

\subsection{Calculation of radial eigenfunctions}

Having determined the values of $l$, we calculate $H_l^{i+\frac12}$ on the cell faces $\rho^{i+\frac12}$, by numerical integration of equation \eqref{eqn:hmode}, subject to the boundary conditions and $H_l^{n_\rho}=0$. The exact scheme used to solve this ordinary differential equation is not important, but a second-order stencil using central differences together with an analytical derivative of $v_{\rm out}$ appears to be adequate. The integration is carried out downward starting from the upper boundary where $H_l=0$, then the whole function is scaled to satisfy the lower boundary condition.

For given $l$, equation \eqref{eqn:gnum} then gives us a simple iterative scheme to determine $G_l^i$ from $H_l^{i+\frac12}$, using the initial value $G_l^0=1$. We observe that this scheme is a discrete approximation of the differential equation  
\begin{equation}
\frac{\partial}{\partial \rho}(G_le^{2\rho})  = \frac{1}{2}l(l+1)\frac{\partial}{\partial \rho}(e^{2\rho})H_l,
\end{equation}
using central differences to approximate the derivatives. This equation in turn follows directly from taking the divergence of \eqref{eqn:beig} for an individual mode.

\subsection{Calculation of expansion coefficients}

The final step is to calculate the expansion coefficients $C_{l,m}$ in \eqref{eqn:dbr}-\eqref{eqn:dbp}, by matching $B_\rho^{0,j,k}$ to an imposed distribution $B_r(s,\phi)$ on the lower boundary. The orthogonality of eigenvectors gives
\begin{eqnarray}
C_{l,m} = \frac{\sum_{j,k}Q_{l,m}^{j+\frac12}\Phi_m^{k+\frac12}B_r^{j+\frac12,k+\frac12}}{\sum_{j,k}\left(Q_{l,m}^{j+\frac12}\Phi_m^{k+\frac12}\right)^2}.
\end{eqnarray}
Care must be taken to ensure that the input data $B_r^{j+\frac12,k+\frac12}$ are flux-balanced, so they are adjusted to have zero sum over the surface.

\section{Comparison Between Potential and Outflow Fields} \label{sec:pf}

\begin{figure}[ht]
\epsscale{0.89}
\plotone{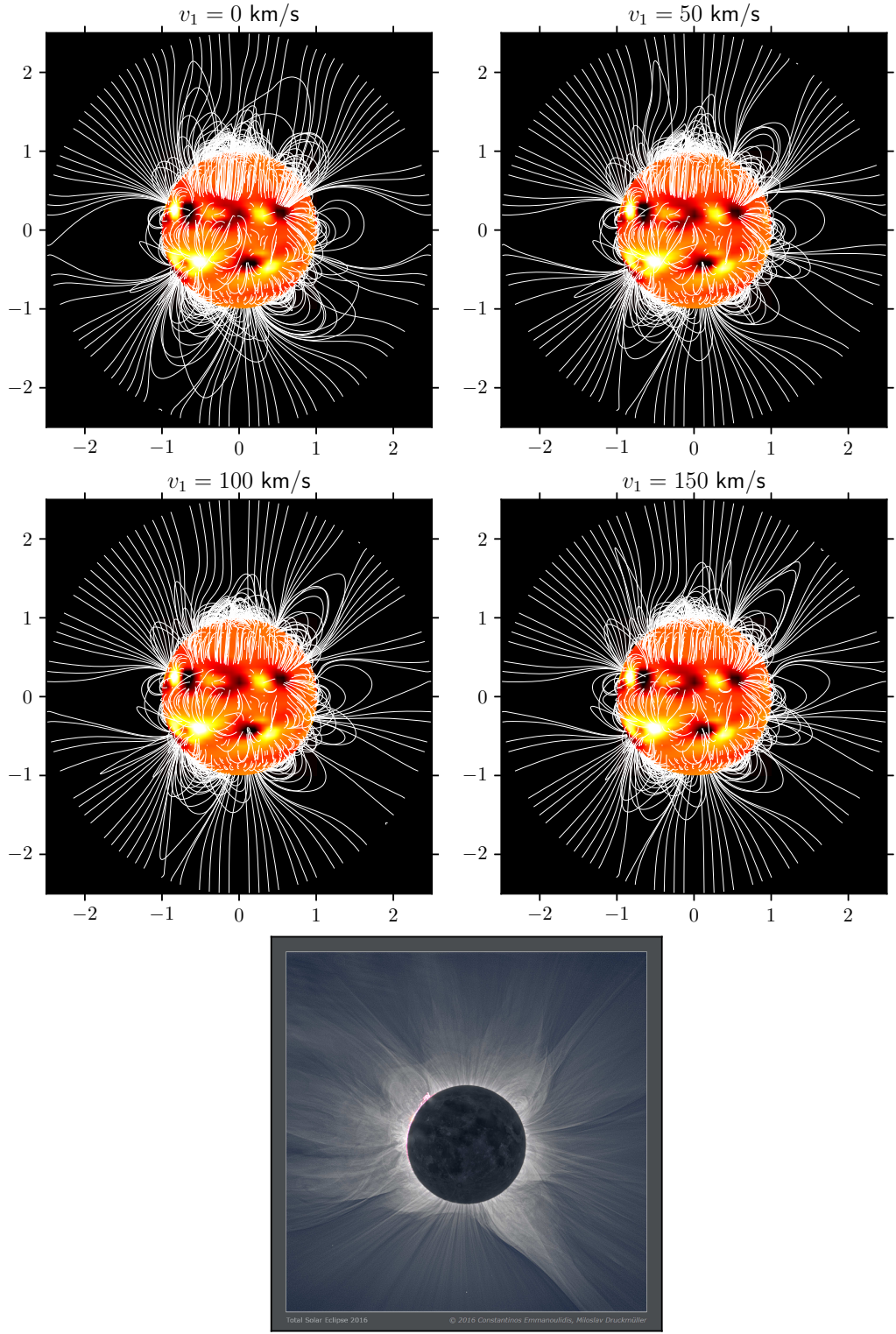}

\caption{Comparison of the magnetic fields for varying outflow velocity $v_1$. The lower boundary data use an HMI synoptic map for Carrington Rotation 2130 so the topology of the corona can be compared to the solar eclipse of the 2016 March 9th. We see a large difference between the potential field (top left) and the outflow fields, illustrated for solar wind speeds up to $150 \,\mathrm{km}\,\mathrm{s^{-1}}$. The photographic image shows a composite of 24 processed eclipse images taken from Tidore, Indonesia on 2016 March 9th (courtesy of C. Emmanoulidis and M. Druckm\"uller, \url{http://www.zam.fme.vutbr.cz/~druck/eclipse/Ecl2016i/Tidore/0-info.htm}).
}
\label{fig:vcomp}

\end{figure}

In this section we discuss the differences between potential fields and the equivalent outflow fields calculated using the method described in this paper. All of the examples in this paper use lower boundary data from the Solar Dynamics Observatory's Helioseismic and Magnetic Imager instrument \citep[HMI,][]{schou2012}. We use the radial component, pole-filled maps in the \texttt{hmi.synoptic\_mr\_polfil\_720s} series \citep{sun2018}.

Figure \ref{fig:vcomp} illustrates the difference in streamer shapes between a potential field and three outflow fields with increasing wind speeds, on the same computational domain with $r_1=2.5R_\odot$. (For all outflow computations in this paper we fix the friction coefficient $\nu_0=5\times 10^{-17}\,\mathrm{s}\,\mathrm{cm}^{-2}$.) For qualitative comparison, Figure \ref{fig:vcomp} includes an observed image of the solar corona taken during the eclipse of 2016 March 9th. 
We observe that close to the solar surface the potential and outflow fields are very similar, but at higher altitudes the solar wind causes quite significant topological changes. In a  potential field, the streamers are petal shaped with a clear boundary layer near $r=r_1$, and all reach exactly to this source surface height. When the solar wind is imposed, it influences the height and shape of the streamers, which begin to change shape at speeds of around $50 \,\mathrm{km}\,\mathrm{s^{-1}}$. At $150 \,\mathrm{km}\,\mathrm{s^{-1}}$ the field lines become radial at a significantly lower altitude than the potential field solution, and there is no boundary layer near $r=r_1$ where the field lines are sharply kinked. The presence of outflow means that closed field lines extend to different heights in different streamers, dependent on the local magnetic field strength. This agrees with coronal observations discussed in \citet{2020ApJ...895..123B}, namely that the coronal field does not become radial at a consistent height and that deviation from the radial direction depends heavily on latitude and the overall activity of the Sun.
It is interesting to note in Figure \ref{fig:vcomp} that the West limb streamers in the outflow fields match more closely than those in the potential field to the eclipse image. At the East limb, the agreement is poorer (for both potential and outflow fields), but direct comparison at this limb is difficult because of the use of a synoptic map for the lower boundary data; longitudes to the east of Central Meridian include ``future'' observations taken after the time of the eclipse.

\begin{figure}[ht]

\plotone{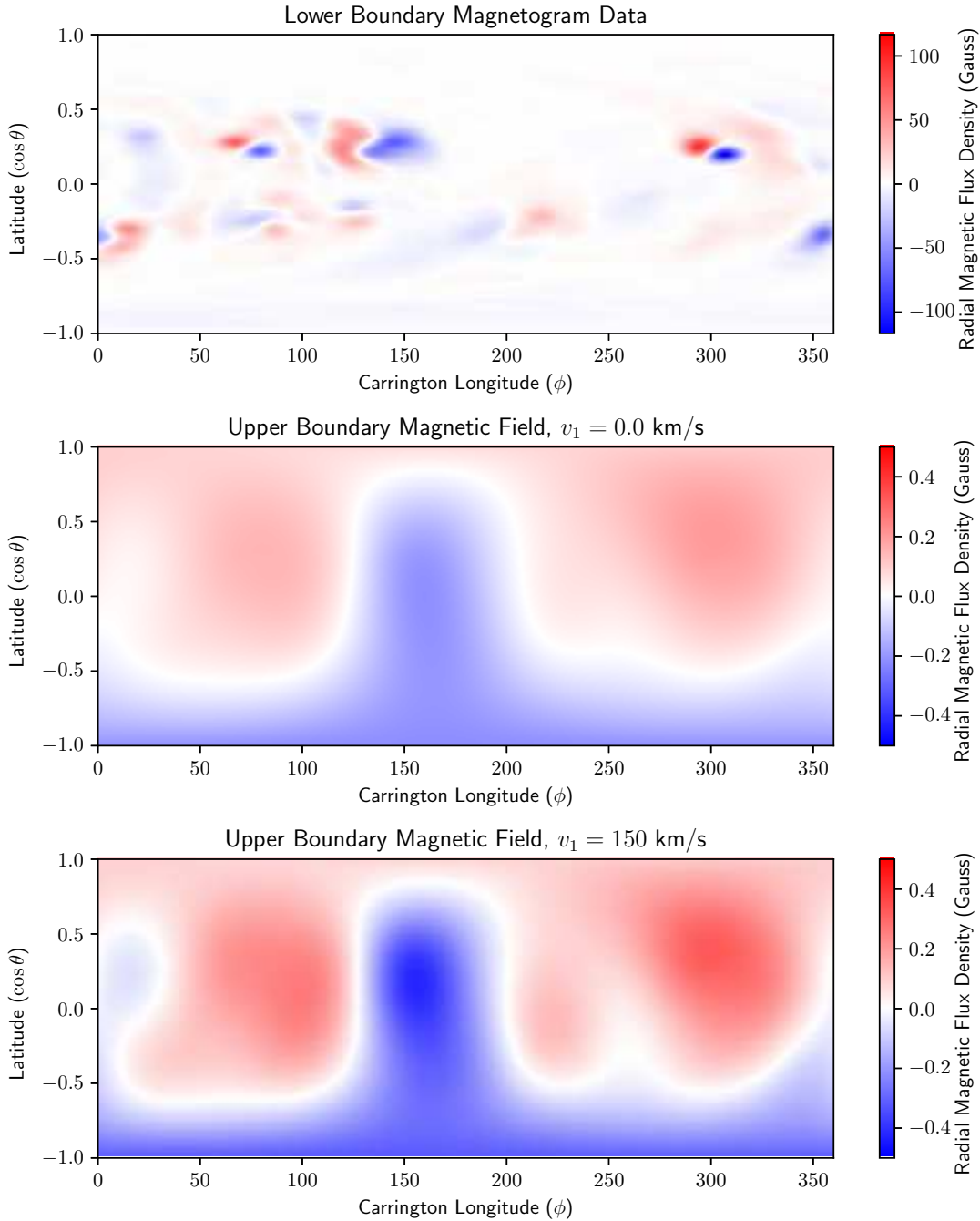} 
\caption{The radial magnetic field at the lower and upper boundaries of the domain, using magnetogram data taken from Carrington Rotation 2165.}

\label{fig:ofluxcomp}
\end{figure}

\begin{figure}[ht]

\plotone{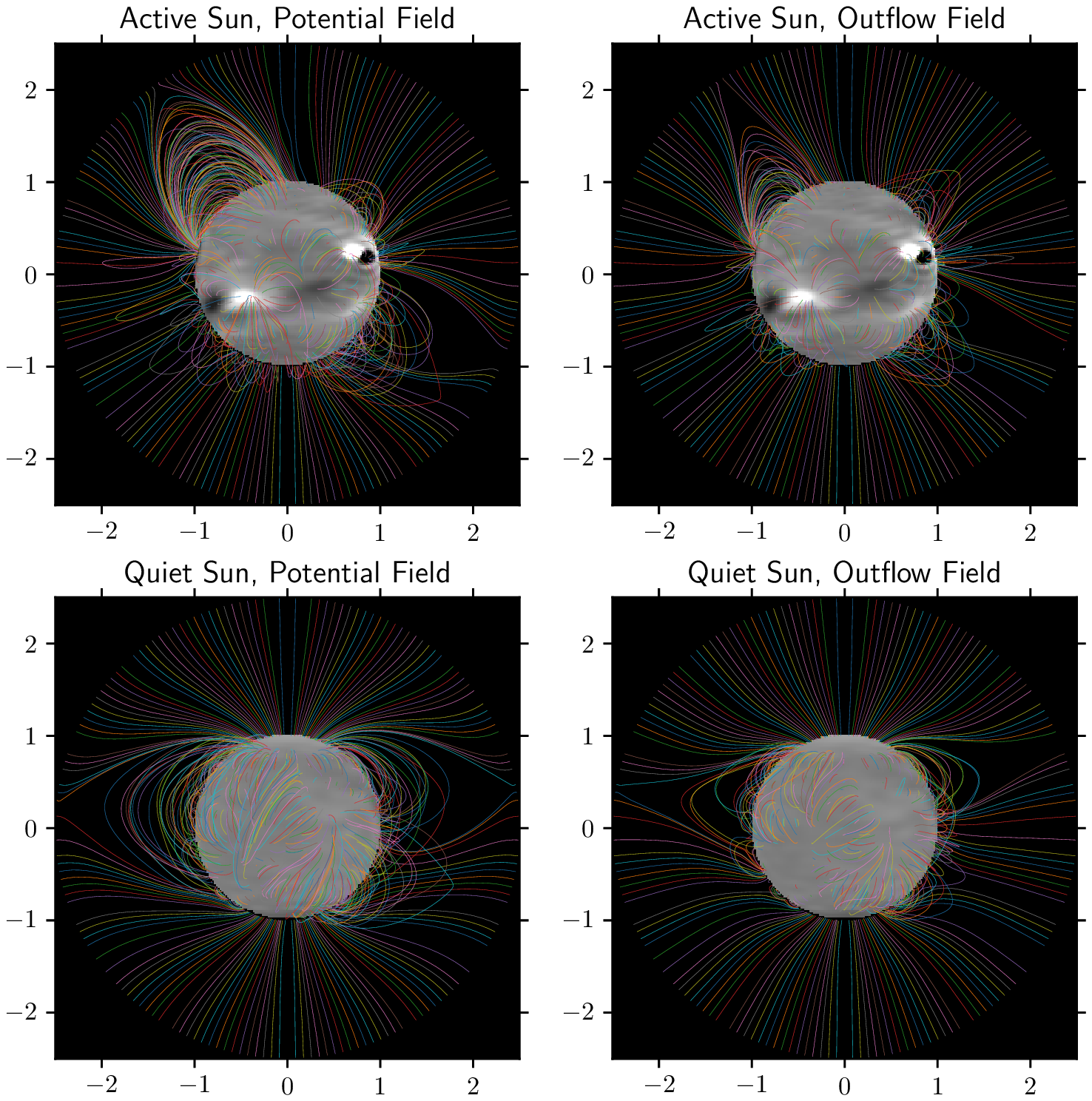}
\caption{Qualitative comparison of potential and outflow fields with $v_1 = 150 \mathrm{km}\,\mathrm{s^{-1}}$. The upper figures show the magnetic field extrapolated from data from Carrington Rotation 2165, when the Sun was relatively active. The lower figures represent a quieter Sun, during Carrington Rotation 2222.}
\label{fig:vcomp2}

\end{figure}

With outflow, the magnetic field is stretched outwards, leading to more open magnetic flux and correspondingly fewer closed field lines within each streamer. The additional open flux is evident in Figure \ref{fig:ofluxcomp}, which compares the radial magnetic field for potential and outflow fields with the same top boundary height, $r_1$. The pattern of positive/negative field polarity at $r=r_1$ is broadly similar in each case and depends only on the magnetic field distribution low in the corona. But, in general, the magnetic field strength high in the corona is larger in the outflow field than in the potential field with the same lower boundary data. Since more of the magnetic flux is open, the closed-field arcades in the outflow field are smaller than if the solar wind is disregarded -- this is clearly seen in Figure \ref{fig:vcomp2}, where we compare the effect of imposed outflow at two stages of the solar cycle, corresponding roughly to solar minimum and maximum. The large closed field regions evident in the potential fields are much smaller in the corresponding outflow solutions, while the magnetic field structure close to the solar surface is little affected.

\begin{figure}[ht]

\plotone{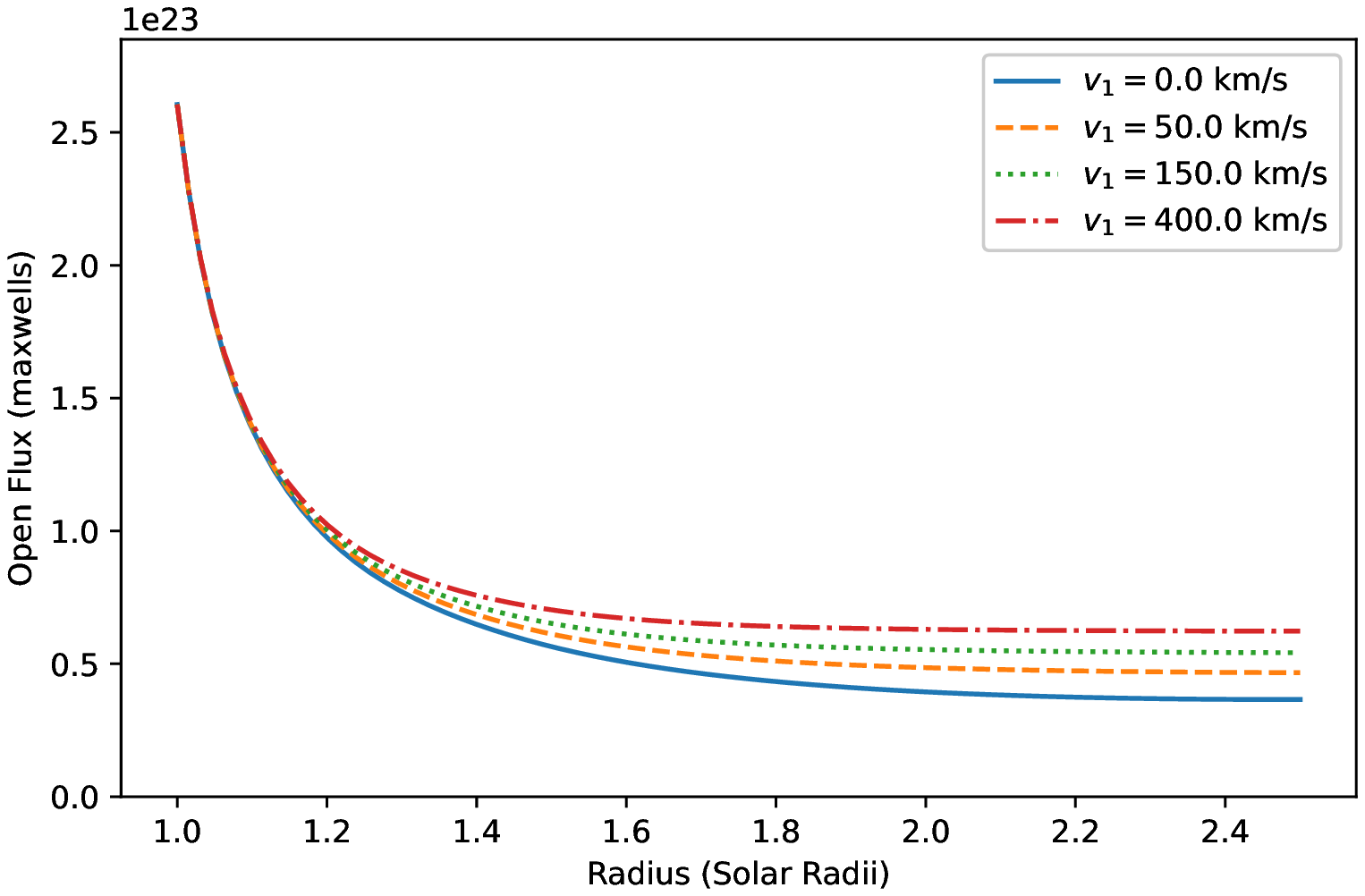} 
\caption{Unsigned open flux as a function of radius for different outflow speeds, for Carrington Rotation 2165.}
\label{fig:radflux}
\end{figure}

A significant difference between the potential and outflow fields is the effect of varying the upper boundary height, $r_1$. For the potential field, increasing $r_1$ will increase the height of the closed field streamers. But in the outflow fields, this height is determined by the outflow velocity rather than the imposed condition of a purely radial magnetic field on $r=r_1$, at least providing that $r_1$ is sufficiently large. With the solar wind model that we have chosen, most streamers extend to less than $2.5R_\odot$, but some extend further. (For comparison with the PFSS model, we set $r_1=2.5R_\odot$ in our computations for this paper.)
To illustrate the behaviour of the outflow fields near to the upper boundary,  Figure \ref{fig:radflux} shows the open magnetic flux in the outflow field as a function of altitude, for solar wind speeds up to $400 \mathrm{km}\,\mathrm{s^{-1}}$ (which is very fast for these altitudes). The open magnetic flux decreases rapidly as we move away from the solar surface, as magnetic field lines curve back towards the sun. The outflow fields exhibit higher flux at larger radii as more of the magnetic field is stretched out by the solar wind. 
We observe that for fast wind speeds, the outflow flux is roughly constant above a radius of $2R_{\odot}$. This is consistent with the observation that the field is roughly radial above this altitude and there are very few closed field lines. It also indicates that the solution is not sensitive to the chosen location of the outer boundary.

\begin{figure}[ht]

\plotone{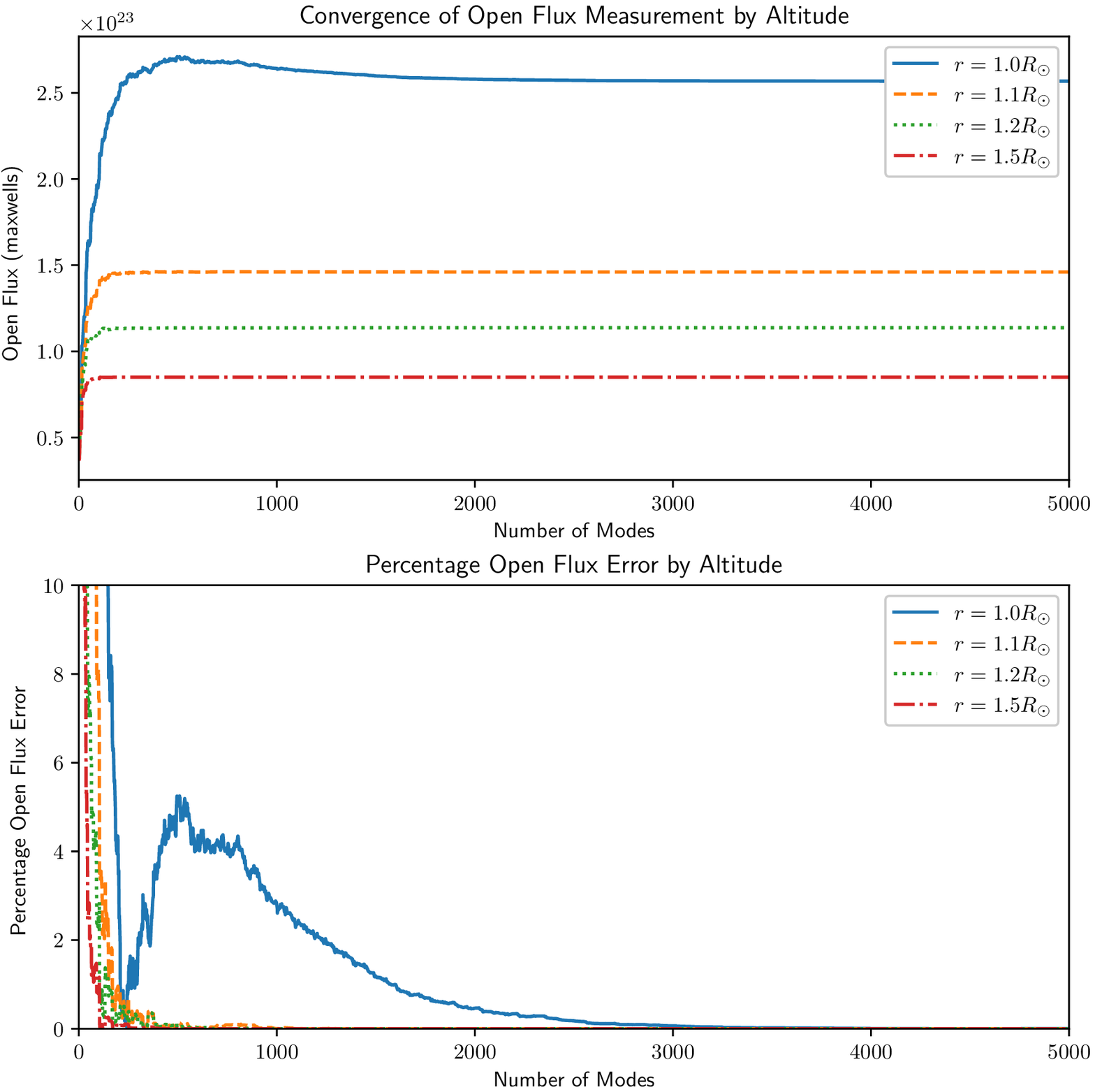} 
\caption{Convergence of the unsigned open flux at different heights, for Carrington Rotation 2165. The top panel shows the computed open flux as more modes are included. The lower-order modes (small $l$ and $m$) are calculated first as in general these contribute more than higher-order modes. The bottom panel shows the corresponding percentage error in the open flux as the number of calculated modes increases. }
\label{fig:converge}
\end{figure}

Finally, we note that the magnetic field at $r=r_1$ is clearly dominated by low-order modes in the azimuthal and latitudinal directions. Thus -- as for the potential field -- it is possible to obtain a close approximation to the true magnetic field at high altitudes while only needing to calculate a relatively small number of modes. This is illustrated in Figure \ref{fig:converge}. The total number of modes (indexed by $l$ and $m$) at this resolution ($180$ x $360$) is $64800$, but we see that the flux measurement converges at all heights within $4000$ modes. Away from the surface, this convergence is even faster - within $1000$ modes. Thus if the region of interest is sufficiently high in the corona, we need only calculate several hundred modes in order to model the magnetic field sufficiently accurately, rather than thousands.
For purposes such as space weather predictions, where the precise magnetic field in the lower corona is unimportant, this saving of computational cost could be useful.

\section{Application to Solar Cycle 24} \label{sec:cycle24}

\begin{figure}[ht]
\plotone{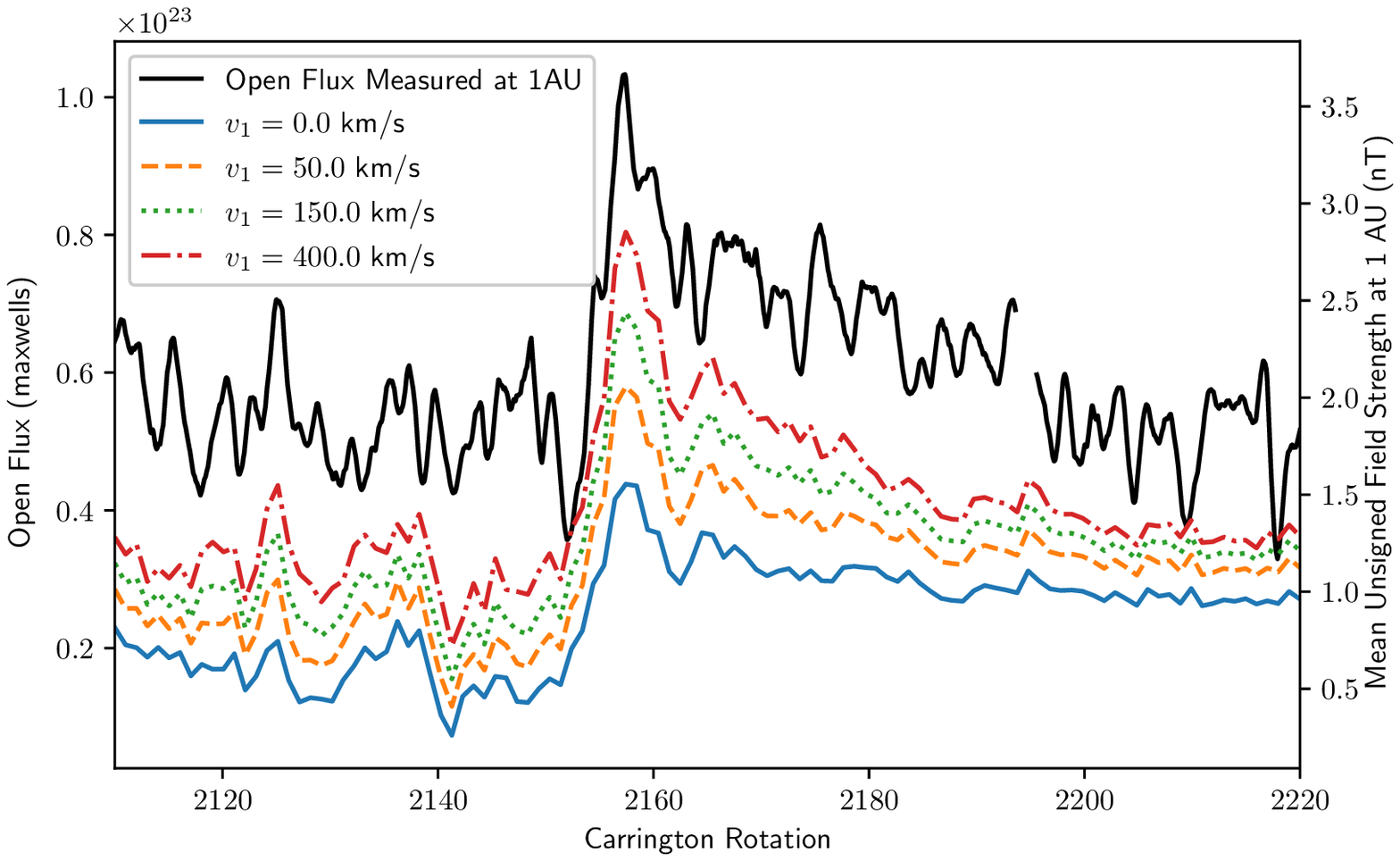} 
\caption{Variation of the open magnetic flux at $r_1=2.5R_{\odot}$ during Solar Cycle 24. The black curve represents OMNI measurements of the magnetic field at 1 AU (smoothed as described in the text), while the dashed, dotted, and dot-dashed curves show predictions from outflow solutions with different speeds.
The lowest curve (blue) shows the result from an equivalent potential field calculated using our model with $v_1=0\,\mathrm{km}\,\mathrm{s}^{-1}$. In all cases, the unsigned open flux is shown on the left axis, while the equivalent field strength at 1 AU (assuming a uniform distribution) is shown on the right axis.}
\label{fig:scycle}
\end{figure}

We now compare the open flux measurements predicted by our outflow model to measurements of the magnetic field at 1 AU extracted from NASA/GSFC's OMNI data set through OMNIWeb. We assume that the total amount of radial magnetic flux at 1 AU is the same as the upper corona and scale the magnetic field strength correspondingly. The data are averaged as in \citet{2010JGRA..115.9112Y} -- namely, an initial daily average of the signed data to smooth out local small-scale fluctuations, then a 27-day running average of the unsigned data for comparison to the global open flux \citep{2009ApJ...700..937L}.
A similar comparison of PFSS extrapolations with these data (up to 2015) is undertaken by \citet{2016ApJ...823...21A}, noting that the definition of open flux in their paper is half the quantity used here.

Figure \ref{fig:scycle} plots the OMNI data against the flux predicted by our model for various solar wind speeds, including the potential field case ($v_1=0$). We observe that throughout Solar Cycle 24 the flux predicted by our potential field model with $r_1=2.5R_\odot$ is consistently an underestimate, as noted in the introduction. The potential field consistently underestimates the measured outflow flux by a factor of more than two, but there is still a strong correlation between the potential field flux and the observations, notably at the large increase around Carrington Rotation 2160. \citet{2016ApJ...823...21A} show that the PFSS open flux can be made to match the observational curve by lowering the source surface height to $r_1\approx 2R_\odot$, although the morphology of the streamers is likely then unrealistic. Those authors also show that a reasonable match to the observed open flux may be obtained with a ``horizontal-current current-sheet source-surface'' (HCCSSS) model. As in our outflow model, the HCCSSS open flux is inflated by the presence of horizontal currents, although unlike in the outflow model the currents flow in the lower part of the domain, and take an arbitrary form that is not directly motivated by observations.

Figure \ref{fig:scycle}  shows that the open flux predicted by our outflow fields also correlates strongly with the OMNI measurements. As discussed in Section \ref{sec:pf}, the outflow fields predict a greater open flux and as such they predict values that more closely match the collected data. Notably, our outflow fields consistently predict more accurate values of the open flux then potential fields, especially for high outflow speeds. It is probable that for sufficiently high outflow speeds the predicted outflow flux would match the OMNI measurements to a high degree of accuracy, but this would likely lead to unrealistic streamer shapes. With a reasonable outflow of $v_1=150\,\mathrm{km}\,\mathrm{s}^{-1}$, about $30-40\%$ of the discrepancy in open flux is accounted for.
It is likely that the remainder must be explained through alternative means. These likely include both steady enhancement from additional low-coronal currents not included here, as well as episodic bursty enhancement from eruptions and coronal mass ejections \citep{2010JGRA..115.9112Y, 2021SoPh..296..109B}. An important further possibility is that the 1 AU data may be overestimating the open flux at $2.5R_\odot$ because some magnetic field lines double back on themselves in the heliosphere \citep{2017JGRA..12210980O}.

\section{Conclusion}

We have described a new method for modelling the global magnetic field in the solar corona. The numerical method is based roughly on existing PFSS models, and in a similar manner requires radial magnetogram data as a lower boundary condition. Our model seeks to improve upon PFSS models by taking into account the effect of the solar wind. We achieve this by seeking equilibrium solutions of the magneto-frictional model, where a radial solar wind outflow function is assumed and specified. Computation times are comparable to PFSS codes, although the methods could be refined further to improve upon this. 
The solutions we find appear more realistic than equivalent potential fields, exhibiting more realistic streamer shapes, reducing the dependence on an arbitrary source-surface height, and increasing the predicted open flux to be closer to OMNI magnetic field measurements throughout Solar Cycle 24. 

Compared to full MHD simulations, our model has the limitation that the solar wind velocity is imposed in a purely phenomenological manner, rather than determined self-consistently as an equilibrium of the full MHD equations. In particular, our method has to rely on several assumptions -- namely that the solar wind velocity is purely radial and only has radial dependence. This is certainly preferable to assuming there is no outflow velocity whatsoever but is still quite a severe limitation. In future, it may be possible to remove these limitations by generalising our method.

Being purely magnetic, however, our method is computationally much less expensive and only requires line-of-sight magnetogram data, as opposed to full vector data and initial conditions for density and pressure/temperature. It thus represents a practical alternative that improves on the commonly-used PFSS model at little extra cost.

For a chosen radial wind speed profile, our solution has a single free parameter: the assumed relaxation rate $\nu_0$. The value for this constant has been determined from experience using the magneto-frictional model but it cannot be calculated directly. Therefore there remains some uncertainly with regards to the most appropriate outflow solution for a given solar wind speed. In future it may be possible to determine $\nu_0$ empirically using the model we have proposed, by comparing streamer shapes to physical observations. In turn, this would then be informative for other magneto-frictional modelling. 

In this paper we have discussed the calculation of a  magnetic field based upon a stretched spherical coordinate system. Altering the differential equations as appropriate could produce an outflow field in standard spherical coordinate system with a similar numerical scheme. We also developed a Cartesian equivalent of the method. As such the outflow fields could be used in place of potential fields in a variety of situations, if so desired.

In conclusion, PFSS fields have been established as a very useful way to model the corona. The ubiquitous use of these fields indicates that computational simplicity is a priority. The methods we present aim to preserve this simplicity. Potential field models are often coupled with current-sheet models to approximate the corona at higher altitudes \citep{2012LRSP....9....6M}. Outflow fields, coupled with accurate functions describing the solar wind velocity at high altitudes, should avoid the need for these extensions, as current sheets between radial magnetic field lines are a natural consequence of our equilibrium solutions. Thus for driving heliospheric models, there is the potential to actually reduce computational complexity by the use of this new method, while simulating the magnetic field more realistically than a PFSS field.

\begin{acknowledgments}
The authors thank UK STFC for supporting this work through a studentship to OEKR and research grant ST/S000321/1 to ARY. The \textit{SDO} data are courtesy of NASA and the \textit{SDO}/HMI science team. We acknowledge use of NASA/GSFC's Space Physics Data Facility's OMNIWeb service, and OMNI data. We also thank Prof. Miloslav Druckmüller of Brno University for the use of his solar eclipse photograph.
\end{acknowledgments}

\bibliography{outflowbib}{}
\bibliographystyle{aasjournal}

\end{document}